\documentclass[pra,twocolumn,showpacs,floatfix,titlepage,superscriptaddress,here]{revtex4-1}

\usepackage{bm}
\usepackage{graphics} 
\usepackage{graphicx}  
\usepackage{latexsym}                % to get LASY symbols
\usepackage{amsmath}     
\usepackage{amsfonts}  
\usepackage{amssymb}
\usepackage{amsthm}
\usepackage{mathrsfs}
\usepackage{dcolumn}
\usepackage{color}
\usepackage{subfigure}

\renewcommand{\phi}{\varphi}
\renewcommand{\>}{\right \rangle}

\newcommand{\ket}[1]{\left |#1\>}

\newcommand{\aver}[1]{\left \langle #1\right \rangle}
\newcommand{\be}{\begin{equation}}
\newcommand{\ee}{\end{equation}}
\newcommand{\bea}{\begin{eqnarray}}
\newcommand{\eea}{\end{eqnarray}}

\begin{document}
\title{Effects of free-electron-laser field fluctuations on the frequency response of driven atomic resonances}
\author{G.~M.~Nikolopoulos}
\affiliation{Institute of Electronic Structure \& Laser, FORTH, P.O.Box 1527, GR-71110 Heraklion, Greece}

\author{P.~Lambropoulos}
\affiliation{Institute of Electronic Structure \& Laser, FORTH, P.O.Box 1527, GR-71110 Heraklion, Greece}
\affiliation{Department of Physics, University of Crete, P.O. Box 2208, GR-71003 Heraklion, Crete, Greece}

\date{\today}

\begin{abstract}
We study the effects of field fluctuations on the total yields of Auger electrons, obtained in the excitation of neutral atoms to a core-excited state by means of short-wavelength free-electron-laser pulses. Beginning with a self-contained analysis of the statistical properties of fluctuating free-electron-laser pulses, we analyse separately and in detail the cases of single and double Auger resonances, focusing on fundamental phenomena such as power broadening and ac Stark (Autler-Townes) splitting. In certain cases, field fluctuations are shown to influence dramatically the frequency response of the resonances, whereas in other cases the signal obtained may convey information about the bandwidth of the radiation as well as  the dipole moment between Auger states.
\end{abstract}


\maketitle

\section{Introduction}
In traditional weak field photo-absorption processes, be it ionization, resonance excitation, scattering, etc., the only property of the source, besides the intensity, that enters the description is the bandwidth. The latter is due to random fluctuations present in any source, and depending on the origin of the radiation, i.e. the nature of the process that gave rise to its production, it may be simply phase fluctuations or in addition intensity fluctuations. But in any case, for single-photon, weak field processes the total yield is sensitive to the bandwidth only, and possibly to the particular form of the source line-shape, depending on the type and details of the experiment.  

The situation changes dramatically, however, when the source is sufficiently intense to induce non-linear processes. Even the simplest case of 2-photon absorption introduces non-trivial departure from the above rule. Thus, the rate of 2-photon ionization, in lowest order perturbation theory (LOPT) is proportional, not simply to the square of the intensity, but rather to the second order intensity correlation function, while N-photon ionization is proportional to the Nth order correlation function \cite{AgaPRA70,LamAAM76}. In yet another example, the strong driving of a single-photon resonant transition between two bound states, by anything other than an idealized monochromatic source, brings into play field correlation functions of all orders \cite{ZolPRA79}. 

The effect of field fluctuations on non-linear photo-interaction processes were discussed extensively during the late 1970’s through the 80’s, which led to the development of a number of theoretical approaches capable of addressing essentially any arrangement of atomic levels, including the continuum, coupled to one or more radiation fields involving fluctuations \cite{AgaPRA78,GeoPRA78,GeoPRA79,ZolPRA79,AgoJPB78,CamPRA93}. The relevant laser sources of that period, although pulsed, were of relatively long duration, in the range of nanoseconds to tens of picoseconds, as a result of which they could to a good approximation be viewed as stationary. Under the assumption of stationarity and ergodicity, certain standard models of stochastic fluctuations developed in quantum optics, such as a field with phase diffusion only, or a chaotic field (in the Glauber sense) \cite{GlaPR63}, were employed in the development of the theories. Even so, the problems involved significant mathematical complexity, but eventually, definitive, clear and in many cases rigorous results were obtained. In addition, a number of seemingly puzzling experimental results were explained.

With the advent of new sources of higher intensity, and much shorter pulse duration, in the few femtosecond range which in the infrared translates into even few cycles, the issue of field fluctuations ceased to be of relevance. That is because such short pulse sources are smooth, Fourier limited, whose bandwidth stems only from the duration, without any fluctuations of either phase or intensity \cite{ScrJPB06}. Although frequency chirping may at times be present and even desirable, it can be reliably considered deterministic, without any stochastic aspects entering the description. 

The situation has changed again with the appearance of the short wavelength, accelerator-based free electron lasers (FELs), with photon energies ranging from the XUV to hard X-rays and pulse durations ranging from a few to a few hundred femtoseconds. Since the frequency in this regime is high, even a few femtoseconds correspond to many cycles of the field. Moreover, although the intensities in terms of W/cm$^2$ may be large, say 10$^{16}$-10$^{18}$, the photon flux and ponderomotive energies are such that perturbation theory of the appropriate order is valid. This means that, depending on the situation, rate equations or density matrix equations within a restricted atomic basis are completely valid \cite{LamPRA11}. Typical cases in point are multiphoton ionization, which can also co-exist with a few-photon resonant excitation of a discrete state, known as REMPI (Resonantly Enhanced Multiphoton Ionization) \cite{LamJPB11,MazJPB12}. But in this range of wavelengths, such discrete states will be unstable, decaying by autoionization or Auger. Also, multiphoton ionization as well as resonant excitation at these large photon energies, involve subvalence electrons; a feature that sets these processes apart from their counterpart in the optical regime, for which the above mentioned techniques, for  treating field fluctuations, were developed.  

But with history repeating itself in some sense, the FEL pulses exhibit significant intensity fluctuations both in frequency and time domain; often referred to as spikes \cite{SalSchYur,Kri06,Ack07}. Given that typically the pulse, although short in terms of femtoseconds, it contains many cycles, and that it also involves amplified spontaneous emission, it could be argued that adopting the model of chaotic radiation should be an adequate approximation.  In many cases and contexts, that may actually be satisfactory. But chaotic radiation, whose electric field classically is described by a complex Gaussian random variable, is a very well defined concept, which is strictly meaningful for a stationary source \cite{GlaPR63}. FEL radiation, however, cannot be assumed to be stationary, while its electric field, depending on the regime of operation, can be far from a Gaussian random variable. The theoretical handling of non-Gaussian random processes represents a formidable challenge. Leaving a more detailed description of the necessary models and  tools for the next section of the paper, it will suffice to point out here that the analytical simplicity, available for a phase diffusion or a chaotic field, is no longer available, which implies a pure numerical simulation of the field and its stochastic properties. Such numerical simulation is not brute force, but based on detailed studies of the properties of the electron beam in the accelerator and the process of lasing taking into account the various regimes of operation of the machines. Our modeling reflects the state of knowledge of the FEL properties as of this time, in the self-amplified-spontaneous-emission (SASE) linear regime of operation. It is expected that with accumulating experience and improving technology, control of the temporal properties of the pulses may improve, with the non-negligible possibility that in the future Fourier limited pulses may be achievable. Until such time, however, the stochastic nature of the radiation needs to be taken into detailed account for its interaction with matter to be fully understood. On the other hand, it should be kept in mind that, the outcome of a non-linear interaction, depending as it does on the correlation functions of the field, provides a probe of the stochastic properties of the source which are very difficult, if possible at all, to probe otherwise. In other words, the fluctuations inherent in the source, which for certain purposes may be a nuisance or dirt effect, from a different perspective contain information on the nature of the process that produced the radiation. That information, which goes well beyond the standard features of intensity and bandwidth, is revealed through the correlation functions. Put otherwise, a non-linear photo-interaction process probes the inner structure of the stochastic process underlying the generation of the radiation.

Although as mentioned above, considerable theoretical know-how, and rules of thumb, concerning the effects of field fluctuations on non-linear processes have accumulated over the years through studies in the optical regime, it is far from obvious that they can be applied intact when it comes to FEL radiation. In fact recent experience in our own work connected to experimental results at FLASH, has led us to the reexamination of the fundamentals of the theory, which has also been the motivation for this paper.  As we show in the sections that follow, some of the accumulated wisdom is to a good approximation transferable to this new regime, while significant modification and reformulation is necessary under certain circumstances. And since our purpose here is to reexamine the fundamentals, we have chosen to focus on two basic themes. One is the simple single-photon excitation of an Auger resonance, the other being double Auger resonance in which one of the fields has the properties of SASE FEL. The arrangement of DR, combined with the strong driving of either of the two transitions, with the other serving as a probe, appears in more than one context, as for example in EIT (Electromagnetically Induced Transparency)  \cite{Eit} or in the strong coupling of two highly excited resonances, embedded in a continuum, which provides an otherwise unavailable test of the theoretical modelling of such resonances \cite{LamPRA81,KarPRL95,TheJPB04,LohCP08}. Our illustration of the effect of fluctuations on these arrangements through  quantitative results and discussions, refer to examples taken from real experimental situations \cite{MazJPB12}, which although specific, serve the purpose of calibration and extrapolation to essentially any analogous arrangement. 

Related studies in the past have not focused on the effect of fluctuations \cite{NikPRA11}, or have been concerned with the energy distribution (spectrum) of the Auger electrons \cite{RohPRA08,KanPRL11}. The aim of the present work is to study the effect of fluctuations (inherently present in the pulses typically produced  in current  SASE FEL sources), on the frequency response of single and double Auger resonances. By contrast to \cite{RohPRA08,KanPRL11}, we do not analyse the actual spectrum of the Auger electrons, but rather we focus on the dependence of the total Auger yield on the detuning of the driving fields from resonance; a less demanding quantity that can be measured in related experiments. The following section is devoted to a brief description of our algorithm and its convergence. The single and double Auger resonances are discussed in Secs. \ref{sec3} and \ref{sec4}, respectively.

\section{Simulation of Chaotic SASE-FEL Pulses}
\label{sec2}
The statistical properties of  light pulses emitted by a SASE-FEL depend crucially on the regime of operation \cite{SalSchYur}. 
In the regime of exponential growth (linear regime), it has been shown, that the pulses exhibit the properties 
of the so-called chaotic polarized  light \cite{SalSchYur,Kri06,Ack07}. This means that the slowly-varying amplitude of 
the electric field is a complex Gaussian random variable, while the instantaneous intensity and the 
energy in the pulses fluctuate according to known distributions. 
As the amplification process reaches saturation and the SASE-FEL enters the non-linear regime, the statistics of the radiation 
deviate significantly from Gaussian \cite{SalSchYur}. The theoretical description of non-Gaussian random processes is a rather difficult task, and to the best of our 
knowledge up to now no analytic results, for the properties of  SASE-FEL radiation in the non-linear regime, have been obtained.  
Finally, a SASE-FEL operating deep in the non-linear regime seems to exhibit properties analogous to chaotic polarized light again \cite{SalSchYur}. 

Over the last decades, a number of FEL simulation codes have been developed, which take into account the details of particular FEL 
facilities \cite{SalSchYur,Pie96Rei99} and are thus capable of simulating its operation throughout the entire regime of parameters. Such codes, however, are rather involved and 
some times adapted to a specific FEL system. On the other hand, for the investigation of  problems pertaining to the interaction of FEL radiation with matter, 
we need more direct ways to simulate the FEL radiation, or at least those properties that are pertinent to the particular process under 
consideration. This is not always an easy task  especially when the radiation involved does not obey Gaussian statistics. 

Throughout this manuscript, we consider light pulses produced from a SASE-FEL operating in the linear regime. The chaotic 
light is a fundamental concept of quantum optics, with the discussion usually limited to stationary and ergodic thermal sources (e.g., see \cite{Loud}). 
In contrast to such a type of sources, SASE-FELs are not continuous sources, and they produce random light pulses which exhibit 
spikes both in time and in frequency domain \cite{SalSchYur,Kri06,Ack07}.  Typically,  the nominal duration of such a pulse is larger than the short time-scale of the 
field fluctuations (i.e., the coherence time).  Such a type of radiation cannot be considered either ergodic or stationary \cite{Good}, and thus simplifications and 
analytic expressions typically used for thermal sources do not apply in the present scenario. For instance, ensemble averages of time-dependent quantities cannot 
be substituted by integrations over time.  In the following we adopt numerical techniques that have 
been developed in the context of quantum optics \cite{FoxPRA88,VanTeiAO80,BilShiPRA90}, in order to produce fluctuating pulses, which exhibit all the properties of  SASE-FEL pulses 
in the linear regime. The details of our algorithm can be found in \cite{VanTeiAO80,BilShiPRA90} but, for the sake of completeness, we briefly summarize here its main aspects. It should also be noted that our algorithm bears analogies to the algorithm used by other authors \cite{RohPRA08}.

\subsection{Algorithm}
\label{sec2a}
In various contexts of physics, one usually deals with noises of different origin, i.e., with various types of fluctuating random variables. Gaussian white noise 
typically refers to a quantity that fluctuates e.g., with time, and it has flat power spectral density (PSD). Formally speaking, it is represented by a random variable 
$\xi(t)$ with Gaussian distributed fluctuations, which satisfies 
\begin{subequations}
\label{gwn}
\bea
\aver{\xi(t)}=\aver{\xi^2(t)}=0,\label{gwn-1}\\
\aver{\xi(t)\xi^\star(t^\prime)}\sim\delta(t-t^\prime),
\label{gwn-2}
\eea      
\end{subequations}
where $\delta(\cdot)$ is the Dirac delta function, while $\aver{\cdot}$ denotes an ensemble (statistical) average. 
Such a noise can be generated e.g., from uniform random variates by means of Box-M\"uller algorithm. 

The generation of Gaussian colored noise, however, is far from straightforward and has been the subject of a rather limited number of papers (e.g., see \cite{FoxPRA88,VanTeiAO80,BilShiPRA90}). 
In all of these papers, the algorithms take as an input uniform or Gaussian white noise, and return Gaussian noise $\zeta(t)$ with 
autocorrelation different from the Dirac delta function i.e., the new variable $\zeta(t)$ satisfies Eq. (\ref{gwn-1}), but 
$\aver{\zeta(t)\zeta^\star(t^\prime)}\sim G_\zeta^{(1)}(t,t^\prime)$, with $ G_\zeta^{(1)}(t,t^\prime)\neq \delta(t-t^\prime)$. The colored-noise algorithm of \cite{FoxPRA88}, relies on the propagation of  a stochastic differential equation, and is limited to the generation of exponentially correlated Gaussian colored noise (Ornstein-Uhlenbeck process). Throughout this work, we adopt the algorithm of \cite{VanTeiAO80,BilShiPRA90}, which is capable of producing Gaussian colored noise for various types of autocorrelation functions, from the corresponding PSD, utilizing the fast Fourier transform. The performance of the algorithm has 
been discussed in the context of particular stochastic differential equations \cite{VanTeiAO80,BilShiPRA90}, and here we will discuss briefly the performance  in the context of simulations pertaining to SASE-FEL pulses.  
  
The algorithm is implemented on a grid of $N_g$ points in frequency domain with step $\delta\omega$, from $\omega_s-\Lambda/2$ to $\omega_s+\Lambda/2$, 
with $\Lambda=N_g\delta\omega$ and $\omega_s$ denoting the central frequency of the PSD  ${\mathscr P}_\zeta(\omega)$ under consideration. An independent complex Gaussian random variable $\xi_k$, corresponds to the $k$th point of the grid (of frequency $\omega_k$) , with $\aver{\xi_k}=\aver{\xi_k^2}=0$ and $\quad \aver{\xi_k\xi_l^\star}=\delta\omega {\mathscr P}_\zeta(\omega_k)\delta_{k,l}$, where $\delta_{k,l}$ is the Kronecker's delta. 
The generation of  noise in the time domain is achieved by means of the discrete Fourier transform,  obtaining complex Gaussian random variables $\zeta(t)$ 
with autocorrelation function $G_\zeta^{(1)}(t,t^\prime)\equiv \aver{\zeta(t)\zeta^{\star}(t^\prime)}$, given by
\begin{subequations}
\bea
G_\zeta^{(1)}(t,t^\prime) = \sum_{k=-N_g/2}^{N_g/2}\delta\omega e^{i\omega_k(t-t^\prime)} {\mathscr P}_\zeta(\omega_k).
\label{cr_zeta}
\eea
It has to be emphasized here that this algorithm generates a stationary random process described by the random variable $\zeta(t)$, with $\aver{\zeta(t)} = 0$ and 
\bea
\aver{|\zeta(t)|^2}=\int_{-\infty}^\infty d\omega {\mathscr P}_\zeta(\omega).
\label{zeta_var}
\eea
\end{subequations}
In the context of SASE-FEL, one is interested in the generation of fluctuating pulses i.e., signals of finite duration and of time-dependent average profile. To this end, the noise produced by means of the aforementioned procedure can be superimposed to a particular profile (envelope)  yielding thus a spiky pulse, analogous to the ones typically produced from SASE FEL sources. Clearly, the new signal cannot be considered  stationary. 

\begin{figure}
\includegraphics[scale=1.0]{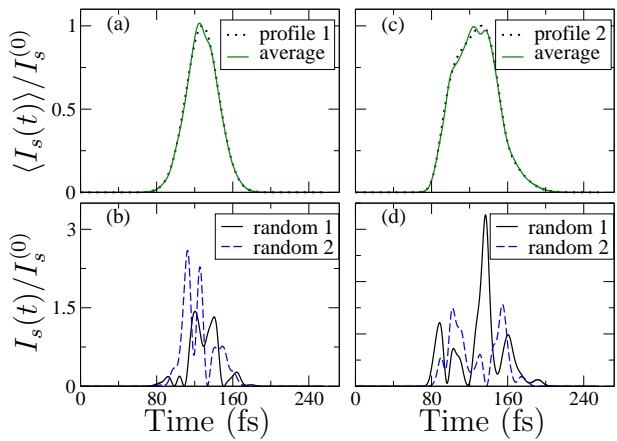}
\caption{(Color online) (a,c) The dotted curves show two of the profiles $f_s(t)$ used in our simulations.  (b,d) A sample of two random spiky pulses, typically produced in a single realization of the algorithm discussed in Sec. \ref{sec2}, by superimposing Gaussian correlated noise ($\sigma_\omega^{-1}=7.2$~fs) with the deterministic profiles of (a,c). The solid curves in (a,c) show the average intensity $\aver{I_s(t)}/I_s^{(0)}$ on a sample of 1000 random spiky pulses.}
\label{fig1}
\end{figure}

\subsection{Application and Convergence}
\label{sec2b}
The above algorithm has been tested for various types of colored noises and various profiles. In the following we present some results pertaining to the generation of fluctuating pulses with Gaussian correlated noise, which is the type of correlations typically observed in different SASE FEL facilities \cite{SalSchYur,Kri06,Ack07,Pie96Rei99,VartPRL11,MitzOE08}.  
Since our noise has to be Gaussian correlated, the algorithm of Sec. \ref{sec2a}, is seeded with a Gaussian PSD for the noise i.e.,  
\bea
{\mathscr P}_\zeta(\omega)= \frac{1}{\sigma_\omega\sqrt{2\pi}}\exp \left [-\frac{(\omega-\omega_s)^2}{2\sigma_\omega^2} \right ],
\label{psd}
\eea 
where $\sigma_\omega$ is the standard deviation of the distribution. As will be discussed later on, the generated noise $\zeta(t)$, is a complex Gaussian random variable (i.e., the real and imaginary parts have Gaussian distributions), and exhibits Gaussian correlations, with the coherence time $T_c$ determined by $\sigma_\omega$. 

To facilitate our theoretical treatment, the amplitude of the electric field with central frequency $\omega_s$ in a single realization of the algorithm (within some non-essential multiplicative constants) is defined as  
\bea
{\cal E}_s(t) = \zeta(t) \sqrt{I_s^{(0)} f_s(t)}, 
\label{Es_t}
\eea
where  $I_s^{(0)} f_s(t)$ is a Fourier-limited (deterministic) pulse profile of finite duration and peak value $I_s^{(0)}$. The intensity of the stochastic pulse in the time domain is simply given by 
\bea
I_s(t)=|{\cal E}_s(t)|^2=I_s^{(0)}f_s(t)|\zeta(t)|^2.
\label{It}
\eea
The deterministic envelope $f_s(t)$ ensures the smooth rise and drop of the intensity, and can be chosen at will. In Fig. \ref{fig1} we show two of the profiles used for 
$f_s(t)$ throughout our simulations, together with a small sample of spiky pulses. The profile of  Fig. \ref{fig1}(a) is a Gaussian  given by 
\bea
f_s(t) = \exp\left[-\frac{(t-t_0)^2}{\tau_s^2}\right ],
\eea
where $\tau_s$ is the pulse duration and $t_0>0$ is the center of the pulse. The profile of Fig. \ref{fig1}(c) is a mathematical construction obtained by superimposing different Gaussians, and enabled us to check the dependence of our results on the shape of the profile. From now on we refer to this profile, as profile 2. In view of Eqs. (\ref{zeta_var}), (\ref{psd}) and (\ref{It}), averaging over a large number of random pulses (which means random functions), one recovers the deterministic profile $f_s(t)$ i.e., 
\bea
\aver{I_s(t)}=I_s^{(0)}f_s(t).
\label{It_ft}
\eea
Typically, one has to average over at least $10^3$ random spiky pulses to ensure a rather good convergence in this respect [see Fig. \ref{fig1}(a,c)]. Throughout our simulations time evolution takes place for $t\in [0, T_f]$, with $T_f$ a finite positive number and the  pulse parameters (center and duration) chosen such that the pulse rises and falls smoothly in this time interval (i.e., formally speaking $I_s(t)\to 0$ as $t\to\{0,T_f\}$).

\subsubsection{Statistics and correlations} 
 
The fluctuations of the instantaneous electric field (not shown here) obey a Gaussian distribution. The probability distribution of the instantaneous intensity, i.e., the probability for an instantaneous measurement to yield a value between $I_s(t)$ and $I_s(t)+{\rm d} I_s(t)$, is described by the negative exponential probability density function (PDF)
\bea
p[I_s(t)] = \frac{1}{\aver{I_s(t)}}\exp\left (-\frac{I_s(t)}{\aver{I_s(t)}}\right ).
\label{ned}
\eea
The energy in a random pulse at a space point in the interaction volume is given by
\bea
W_s \propto \int_0^\infty I_s(t) \rm{d}t,  
\eea
and it fluctuates from pulse to pulse. The corresponding probability distribution is the Gamma PDF  given by 
\bea
p(W_s) = \frac{M^MW_s^{M-1}}{\Gamma(M)\aver{W_s}^{M}} \exp\left (-M\frac{W_s}{\aver{W_s}} \right ),  
\label{gd}
\eea
where $\Gamma(M)$ here is the Gamma function with argument $M=\aver{W_s}^2/\aver{(W_s-\aver{W_s})^2}$. 
The parameter $M$ represents the average number of modes in a radiation pulse.  
All of the above properties are in agreement with experimental observations and theoretical results pertaining to various SASE-FEL facilities \cite{SalSchYur,Kri06,Ack07,Pie96Rei99}.  Although, owing to the inevitable focusing in experiments with high intensity sources, the interaction volume relevant to a particular experiment needs to be taken into account in the quantitative interpretation of data, this is a separate issue. The temporal characteristics of the radiation are, however, the same throughout the interaction volume, which means that our results can be applied to any experimental arrangement by folding in the geometry of the interaction volume (see also related remarks in Sec. \ref{sec5}).

The negative exponential distribution for the intensity and the Gamma distribution for the energy  are intimately connected to a field that obeys Gaussian statistics \cite{Loud,Good}. 
The corresponding first-order coherence, however, is not related to the Gaussian statistics of the field, but rather to the chosen PSD of the generated noise.  Choosing a Gaussian PSD of the form (\ref{psd}), one expects a Gaussian first-order autocorrelation function.  By definition $G^{(1)}(t,t^\prime)=\aver{{\cal E}_s(t){\cal E}_s^\star(t^\prime)}$, and 
using Eqs. (\ref{Es_t}) and (\ref{It_ft}) we obtain
\bea
|G^{(1)}(t,t^\prime)|&=&\sqrt{\aver{I_s(t)}\aver{I_s(t^\prime)}} |G_\zeta^{(1)}(t,t^\prime)| \nonumber\\
&=&I_s^{(0)}\sqrt{f_s(t)f_s(t^\prime)} |G_\zeta^{(1)}(t,t^\prime)|,
\label{G1}
\eea
where $G_\zeta^{(1)}(t,t^\prime)=\aver{\zeta(t)\zeta^\star(t^\prime)}$. We see therefore that the first-order autocorrelation function for the field not only depends on  
the statistical properties of the noise $\zeta(t)$, but also on the ensemble average of SASE-FEL pulses $\aver{I_s(t)}$. Only in the case of stationary fields one has constant $\aver{I_s(t)}$, and thus $G^{(1)}(t,t^\prime)$ depends solely on the noise correlations and not on the average profiles (we return to this point at the end of the section). The modulus of the degree of first-order temporal coherence is defined as \cite{Loud} 
\bea
|g^{(1)}(t,t^\prime)|=\frac{|G^{(1)}(t,t^\prime)|}{\sqrt{\aver{I_s(t)}\aver{I_s(t^\prime)}}}. 
\eea
and is equal to $|G_\zeta^{(1)}(t,t^\prime)|$. For the Gaussian PSD (\ref{psd}), we have 
\bea
|g^{(1)}(t,t^\prime)| = \exp\left [ -\frac{\sigma_{\omega}^2(t-t^\prime)^2}{2}\right ] = |G_\zeta^{(1)}(t,t^\prime)|,
\label{Gcr}
\eea
with $\sigma_{\omega}$ determining how fast 
the correlations drop with the delay.

For a field that obeys Gaussian statistics, Wick's theorem implies that higher-order degrees of coherence can be expressed in terms of $g^{(1)}(t,t^\prime)$, and hence they are determined by the spectral properties of the field \cite{Loud}.  Throughout our simulations we have used various types of ${\mathscr P}_\zeta(\omega)$ and thus of correlations, some of which are summarized in table \ref{tab1}, together with the corresponding bandwidths $\gamma$ [i.e., the FWHM of ${\mathscr P}_\zeta(\omega)$].  
The coherence time is defined as 
\bea
T_c \equiv \int_{-\infty}^{\infty} |g^{(1)}(v)|^2 dv,
\eea
where $v=t-t^\prime$, and is also shown in table \ref{tab1}. For the discussion to follow, it is worth keeping in mind that for fixed $\sigma_\omega$ and for $\tilde{v}<2$, the exponential $|g^{(1)}(\tilde{v})|$ exhibits the fastest drop, and the hyperbolic secant the slowest one.  

\begin{table*}\caption{Various PSDs with the corresponding degrees of first order temporal coherence, bandwidths, and coherence times.} \label{tab1}
\begin{ruledtabular}
\begin{tabular}{lcccc}
Label & ${\mathscr P}_\zeta(\tilde{\omega})$ \footnotemark[1] &  $[|g^{(1)}(\tilde{v})|]$\footnotemark[2] & $\gamma$ & $T_c$\\
\hline
Exp. &  $ [\sigma_\omega\pi(\tilde{\omega}^2+1)]^{-1}$  &   $\exp[-\tilde{ v}]$ & $2\sigma_\omega$ & $\sigma_\omega^{-1}$\\ 
Gauss. &   $\exp[-{\tilde{\omega}^2 }/2]/(\sigma_\omega\sqrt{2\pi})$ &   $\exp[-\tilde{v}^2/2] $ & $2\sigma_\omega\sqrt{2\ln(2)}$ & $\sqrt{\pi}\sigma_\omega^{-1}$\\ 
Sech. &  $ {\rm sech}(\pi \tilde{\omega})/\sigma_\omega$  &  ${\rm sech}(\tilde{v}/2)$  & $2\sigma_\omega{\rm cosh}^{-1}(2)/\pi$ & $4\sigma_\omega^{-1}$\\ 
\end{tabular}
\end{ruledtabular}
\footnotetext[1]{$\tilde{\omega} = \omega/\sigma_\omega$}
\footnotetext[2]{$\tilde{v} = (t-t^\prime)\sigma_\omega$}
\end{table*}

\subsubsection{Bandwidth of fluctuating pulses}

The bandwidth $\gamma$ is determined by the noise in the pulse, and a question arises here as to the bandwidth of the fluctuating pulses produced by superimposing this random noise with the Fourier-limited average profile  $\aver{I_s(t)}$. This profile ensures the smooth rise and drop of the intensity for $t\in[0,\infty)$ and thus we can safely assume that $I_s(t)$ is a square integrable function. One can then  define the energy spectral density of the random pulses as  ${\mathscr E}_s(\omega)=\aver{|{\cal E}_s(\omega)|^2}$, where ${\cal E}_s(\omega)$ is the Fourier transform of ${\cal E}_s(t)$, and using  Eqs. (\ref{Es_t}) and (\ref{It_ft}) we obtain 
\begin{widetext}
\bea
{\mathscr E}_s(\omega) = \frac{1}{(2\pi)^2}\int_{-\infty}^{\infty}\int_{-\infty}^{\infty}  dt dt^\prime \sqrt{\aver{I_s(t)}\aver{I_s(t^\prime)}}G_\zeta^{(1)}(t,t^\prime) e^{i\omega(t-t^\prime)}.
\eea
\end{widetext}
This is the Wiener-Khintchine theorem for non-stationary square integrable signals. For given $\aver{I_s(t)}$ and specified noise, one can obtain ${\mathscr E}_s(\omega)$, and thus the associated bandwidth. Usually one is interested in the normalized spectrum $\tilde{\mathscr E}_s(\omega)$, which is obtained by dividing ${\mathscr E}_s(\omega) $ by $\int_{-\infty}^{\infty} {\mathscr E}_s(\omega)d\omega$. 

\begin{subequations}
For Gaussian $\aver{I_s(t)}$ of duration $\tau_s$, and the Gaussian correlated noise of  Eq. (\ref{Gcr}), $\tilde{\mathscr E}_s(\omega)$ is also Gaussian. We obtain  
\bea
\tilde{\mathscr E}_s(\omega) = \frac{2\sqrt{\ln(2)} }{\sqrt{\pi}\Delta\omega_s}\exp\left [-\frac{4\ln(2)\omega^2}{\Delta\omega_s^2}\right ],
\eea
where  $\Delta\omega_s$ is the bandwidth (FWHM) of $\tilde{\mathscr E}_s(\omega)$ given by 
\bea
\Delta\omega_s = \frac{2\sqrt{1+2\tau_s^2\sigma_\omega^2}\sqrt{\ln(2)}}{\tau_s}.
\label{bw1}
\eea

The case of Fourier-limited pulses corresponds to $\sigma_\omega\tau_s\ll 1$, with the result 
\bea
\Delta\omega_s^{\min}  = \frac{2\sqrt{\ln(2)}}{\tau_s} =  \frac{4\ln(2)}{\Delta t_s},
\label{fl_bw}
\eea
which is the time-bandwidth relation for Gaussian Fourier-limited pulses usually found in standard textbooks \cite{Yariv}, 
with the FWHM of $\aver{I_s(t)}$ denoted by $\Delta t_s = 2\sqrt{\ln(2)}\tau_s$. Hence, Eq. (\ref{bw1}) may be also expressed as
\bea
\Delta\omega_s = \Delta\omega_s^{\min} \sqrt{1+2\chi^2},
\label{bw1c}
\eea
where 
\bea
\chi \equiv \sigma_\omega\tau_s.
\label{chi_eq}
\eea
\end{subequations}
In the other limit, even for $\chi>2$, one essentially 
approaches the limit of stationary source with $\Delta\omega_s\simeq \gamma$, where $\gamma$ is given in table \ref{tab1}. As mentioned above, $\sigma_\omega$ is inversely proportional to the coherence time of the noise 
entering the pulse (e.g., see table \ref{tab1}), and thus $\chi$ characterizes the number of spikes in a Gaussian  pulse profile 
of duration $\tau_s$.  The derivation of simple analytic expressions for ${\mathscr E}_s(\omega)$  is a rather difficult task, when the noise is not Gaussian correlated and/or for arbitrary average profiles. 
In any case, however, the  bandwidth of a fluctuating pulse $\Delta\omega_s$ is expected to be determined by the bandwidth of the Fourier-limited average  pulse profile $\aver{I_s(t)}$ and the product $\chi$, albeit perhaps through a rather complicated expression.  From another point of view, expressing $\sigma_\omega$ in terms of $\gamma$ (see table \ref{tab1}), Eq. (\ref{bw1c}) reads  
\bea
\Delta\omega_s = \Delta\omega_s^{\min}\sqrt{1+\left (\frac{\gamma}{\Delta\omega_s^{\min}}\right )^2}.
\label{bw2}
\eea
This relation shows that for pulses with Gaussian average profile that exhibit Gaussian-correlated fluctuations, the bandwidth is 
the geometric mean of the bandwidths corresponding to the Fourier-limited average profile and the fluctuations. For $\gamma\gg \Delta\omega_s^{\min}$, the bandwidth of the pulse is fully determined by the fluctuations i.e., $\Delta\omega_s\simeq\gamma$, 
whereas  for $\gamma\ll \Delta\omega_s^{\min}$, we have the case of a Fourier-limited pulse with $\Delta\omega_s\simeq\Delta\omega_s^{\min}$.

\begin{figure}
\includegraphics[scale=1.0]{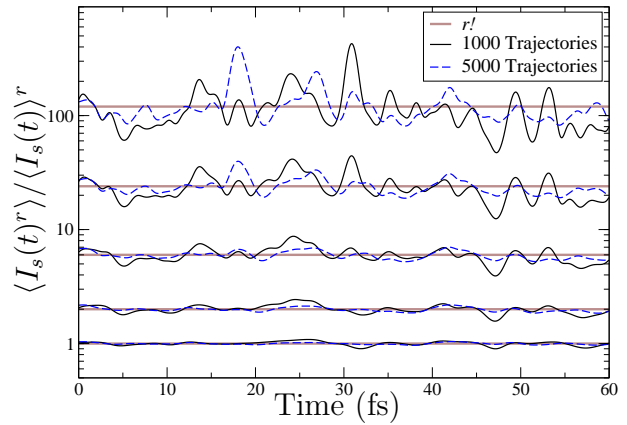}
\caption{(Color online) Convergence of the algorithm discussed in Sec. \ref{sec2}. The ratio $\aver{I_s(t)^r}/\aver{I_s(t)}^r$ is plotted as a function of time, with the average taken over two different numbers of pulses (realizations). The thick grey (horizontal) lines refer to $r!$, which is the theoretically expected value of the ratio. Other parameters: $\tau_s=10$ fs, $\sigma_{\omega}^{-1}=2$ fs.
}
\label{fig5}
\end{figure}

\subsubsection{Convergence with respect to intensity moments}
Let us discuss briefly the convergence of the algorithm in terms of the moments of 
the instantaneous intensity. In view of Eq. (\ref{ned}) we have, 
\bea
\aver{I_s(t)^r}\equiv \int_0^\infty I_s(t)^r p[I_s(t)] \textrm{d}I_s(t)=r! \aver{I_s(t)}^r.  
\label{I_to_r}
\eea
As depicted in Fig.  \ref{fig5}, the ratio  $\aver{I_s(t)^r}/\aver{I_s(t)}^r$ fluctuates around $r!$, with decreasing amplitude of oscillations 
as we average over more pulses (realizations) \cite{remark1}. For increasing $r$, the convergence is slower, 
in the sense that more realizations are needed to approximate adequately $r!$. Note, however, that even for $10^3$ or $5\times 10^3$ realizations, the amplitude 
of the oscillations for $r\leq 5$ is smaller than the gap between successive factorials, and thus one can distinguish unambiguously between different values of $r$ in this regime. 
The appearance of the $r!$ factor in Eq. (\ref{I_to_r}) is a distinctive feature of chaotic light \cite{Loud}, which is expected to affect considerably fundamental nonlinear physical processes, such as the multiphoton atomic ionization \cite{LamPRA11}.  

\subsubsection{Stationarity}
The aforementioned algorithm can be used in studies pertaining to interactions between matter and SASE-FEL radiation with the  statistical properties described above, and various types of autocorrelation functions. In the following we explore the influence of fluctuations on the frequency response of atoms. Analogous studies in the past have focused on stationary fields i.e., for time-independent $\aver{I_s(t)}$ \cite{GeoPRA79,ZolPRA79,CamPRA93}, while most of them considered exponentially-correlated fluctuations. In the present SASE-FEL facilities such conditions are not satisfied, and the ensemble average intensity $\aver{I_s(t)}$ has a finite duration $\tau_s$. It may be possible to assume some sort of stationarity only if the coherence time $T_c$ is much smaller than the pulse duration $\tau_s$, and the field is observed for a time window much shorter than $\tau_s$. Such a condition may be fulfilled in practise when one monitors directly the light, but it is hard to be fulfilled when matter interacts with SASE-FEL pulses. Typically, in such cases the target (atoms, molecules, etc) experience the rise as well as the fall of each one of the pulses, and one simply monitors the products of the interaction (i.e., electrons, ions, etc).  Under such circumstances, the SASE-FEL radiation may or may not be considered stationary and any theoretical description has to take into account the finite temporal width of $\aver{I_s(t)}$.  As long as $\aver{I_s(t)}$ is a smooth function, its  details besides the duration $\tau_s$, are not expected to play a significant role in most cases. For the problems under consideration, we have confirmed this fact by considering different profiles for $\aver{I_s(t)}$, two of which are shown in Figs. \ref{fig1}(a) and (c).

\begin{figure}
\begin{center}
\includegraphics*[width=80mm]{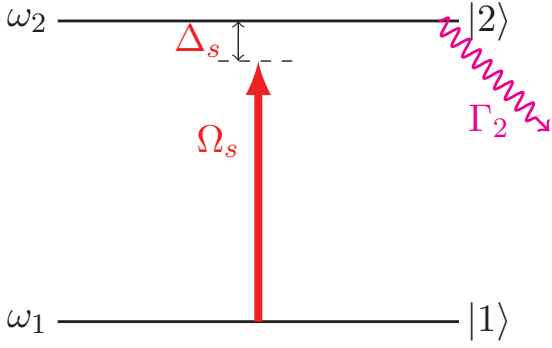}% Here is how to import EPS art
\caption{\label{tls_fig} (Color online) An atomic transition pertaining to a sub-valence electron, driven by chaotic SASE-FEL pulses. The resulting inner-shell vacancy in the excited atom 
decays with rate $\Gamma_2$.}
\end{center}
\end{figure}

\section{Single Auger resonance}
\label{sec3}
Consider the case of a sub-valence atomic transition depicted in Fig. \ref{tls_fig}. A SASE-FEL beam is focused on a target of neutral atoms, inducing an electric-dipole transition of an inner-shell electron from state $\ket{1}$ to a highly excited state $\ket{2}$. The relaxation of the sub-valence vacancy in the excited atom via an Auger decay, gives rise to Auger electrons which are observed 
in the experiment.  Let $\hbar\omega_j$ denote the energy of the state $\ket{j}$, and let $\Gamma_2$ be the rate associated with the Auger decay. 
The electric field of the radiation is given by
\bea
{\bf E}_s(t)=\left [{\cal E}_s(t)e^{i\omega_st}+{\cal E}_s^\star (t)e^{-i\omega_st}\right ] {\bf e}_s,
\eea
where $\omega_{s}$ denotes the central frequency of the spectrum, ${\bf e}_s$ is the polarization vector, and ${\cal E}_s(t)$ is the fluctuating complex amplitude. 

\subsection{Formalism}
\label{sec3a}
Throughout this work ${\cal E}_s(t)$ is treated as a stochastic complex Gaussian random function and is generated along the lines of the previous section. 
The instantaneous Rabi frequency $\Omega_{s}(t)$, given by 
\bea
\Omega_{s}(t)=\frac{\vec{\mu}_{12}\cdot{\bf e}_s {\cal E}_s(t)}{\hbar},
\label{stocOmega0}
\eea 
is also a stochastic complex Gaussian random function with zero mean and variance determined by the 
variance of the field. In this definition, $\hat{\vec{\mu}}$ is the electric dipole operator, and $\vec{\mu}_{12}$ is the transition dipole moment for $\ket{1}\leftrightarrow\ket{2}$. In the following for the sake of brevity we also write $\mu_{12}=\vec{\mu}_{12}\cdot{\bf e}_{s}$. 
In view of Eqs. (\ref{Es_t}) and (\ref{It_ft}), we have
\bea
\Omega_s(t) = \frac{\mu_{12}\sqrt{\aver{I_s(t)}}}{\hbar}\zeta(t)= \Omega_s^{(0)}  \sqrt{f_s(t)} \zeta(t).
\label{stocOmega1}
\eea
where 
\bea
\Omega_s^{(0)} =\frac{\mu_{12}\sqrt{I_s^{(0)}}}{\hbar}
\label{stocOmega2}
\eea
is the peak value of the Rabi frequency. 

The problem can be formulated in the framework of the reduced atomic density matrix with elements $\rho_{ij}(t)$. 
In the rotating-wave approximation, the equations of motion for $\rho_{ij}$ read  
\begin{subequations}
\label{tla_ode}
\bea
&&\frac{\partial\sigma_{11}}{\partial t}=2{\rm Im}[\Omega_s^\star\sigma_{12}]\\
&&\frac{\partial\sigma_{22}}{\partial t}=-\Gamma_2\sigma_{22}-2{\rm Im}\left [\Omega_s^\star\sigma_{12} \right ]\\
&&\frac{\partial\sigma_{12}}{\partial t}=\left (i\Delta_s-\frac{\Gamma_{21}}{2}\right )\sigma_{12}+i\Omega_s(\sigma_{22}-\sigma_{11}),
\eea
where $\rho_{ii}=\sigma_{ii}$, $\rho_{12}=\sigma_{12}e^{i\omega_s t}$, $\Delta_s=\omega_{21}-\omega_{s}$ is the detuning of the field from resonance, while Stark shifts have been neglected. In the absence of other types of (in)homogeneous broadening mechanisms, we have $\Gamma_{12}=\Gamma_2$. In the presence of fluctuations in the electric field, this is a set of coupled stochastic differential equations.
 
 The frequency response of the atoms to the SASE-FEL radiation, as we vary $\omega_s$ around resonance, is obtained by monitoring the Auger electrons. The total probability for Auger decay is given by 
\bea
Q_2 = \Gamma_2\int_0^{\infty}\sigma_{22}(t)\,{\rm d}t.
\label{q2_eqa}
\eea
Alternatively, we can add to the equations of motion for $\sigma_{ij}$, the following differential equation 
\bea
\frac{\partial Q_2}{\partial t}=\Gamma_2\sigma_{22},
\label{q2_eqb}
\eea
\end{subequations}
where $\Gamma_2$ is the probability per unit time for Auger decay. 

In each realization of our simulations a SASE FEL pulse ${\cal E}_s(t)$ is generated randomly according to the algorithm described in Sec. \ref{sec2}. Subsequently, Eqs. (\ref{tla_ode}) are  propagated from $t=0$ to $t=T_f$, keeping track of $Q_2$ at $t=T_f$, where $T_f\gg\tau_s$; which is essentially equivalent to taking the upper limit of Eq. (\ref{q2_eqa}) to infinity. Averaging over a large number of random pulses, we obtain the average stochastic signal $\aver{Q_2}$. 

In connection to the scheme of Fig. \ref{tls_fig}, it is worth mentioning here the recent experiment by Mazza {\em et al.} \cite{MazJPB12}, pertaining to the spectral response of the  
Auger resonance 3d$\to$5p in Kr. This experiment was in the regime of weak fields and long pulses (with respect to $\Gamma_2^{-1}$). By contrast, the pulse durations considered 
throughout the present work are comparable to $\Gamma_2^{-1}$, while a broad range of peak values for the Rabi frequency is considered. To keep our formalism as general as possible, 
in the following discussion the various quantities are measured in units of the natural linewidth $\Gamma_2$. The actual value of $\Gamma_2$ depends on the particular atomic resonance under considerations. For instance, the linewidth for the Auger resonance 3d$\to$5p in Kr is $83$ meV, which corresponds to a lifetime of about $8$ fs \cite{MazJPB12}. The Rabi frequencies we consider here for the FEL radiation vary from 0.1 to  2$\Gamma_2$, corresponding to intensities $10^{14}-10^{17}$W/cm$^{2}$ for the 3d$\to$5p resonance in Kr, whereas the intensities in the experiment of \cite{MazJPB12} were much lower.
 
\begin{figure}
\begin{center}
\includegraphics[width=62mm, angle=270]{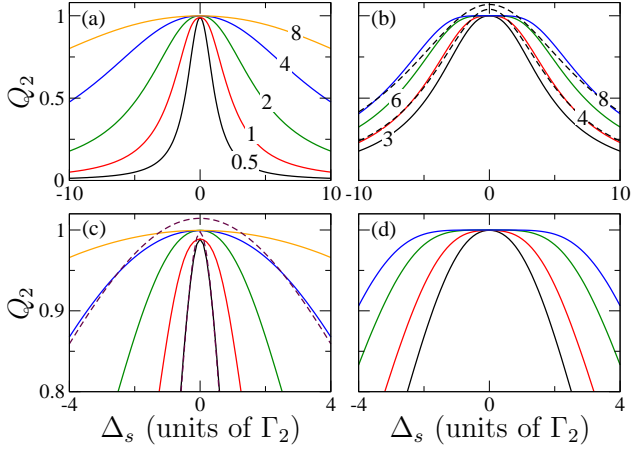}% Here is how to import EPS art
\caption{\label{linedet} (Color online) Single resonance driven by a Fourier-limited pulse. The Auger signal $Q_2$ is plotted as a function of the detuning $\Delta_s$. (a) Various  values of the ratio $\Omega_s^{(0)}/\Gamma_2$ and for $\tau_s = 3\Gamma_2^{-1}$;  (b) Various values of $\Gamma_2\tau_s$ for fixed $\Omega_s^{(0)}=2\Gamma_2$. A close-up of (a) and (b) around the peak is shown in (c) and (d), respectively. The dashed curves show the best Lorentzian fits to the numerical data for (b) $\Gamma_2\tau_s=4$ and $\Gamma_2\tau_s=8$; (c)  $\Omega_s^{(0)}=0.5\Gamma_2$ and $\Omega_s^{(0)}=4.0\Gamma_2$. Other parameters: Gaussian pulse profile, $T_f=32\Gamma_2^{-1}$.}
\end{center}
\end{figure}

\begin{figure}
\begin{center}
\includegraphics[width=83mm]{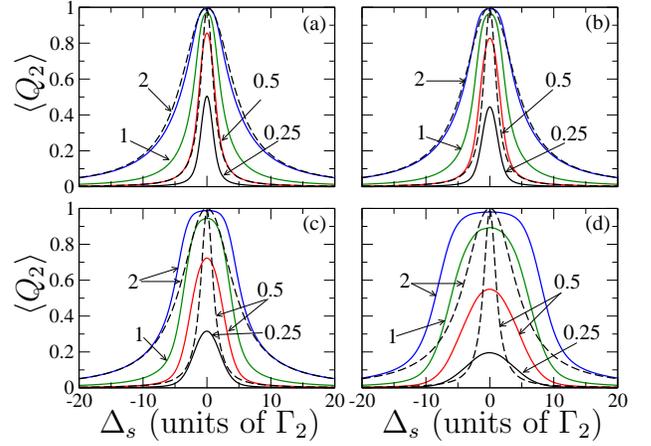}% Here is how to import EPS art
\caption{\label{linerand} (Color online) Single resonance driven by stochastic pulses. Stochastic Auger signal averaged over 5000 stochastic pulses as a function of the detuning $\Delta_s$, for various  values of the ratio $\Omega_s^{(0)}/\Gamma_2$ and (a) $\chi=1.67$;  (b) $\chi=2.5$; (c) $\chi=5.0$;  (d) $\chi=10.0$.  
For the sake of comparison, the signal for a Fourier-limited Gaussian pulse with intensity profile the same as $\aver{I_s(t)}$ is also shown for $\Omega_s^{(0)}=0.5\Gamma_2$  and $\Omega_s^{(0)}=2\Gamma_2$ (dashed curves). Other parameters: Gaussian profile $\aver{I_s(t)}$  with $\tau_s = 3\Gamma_2^{-1}$, $T_f=32\Gamma_2^{-1}$, Gaussian-correlated noise.}
\end{center}
\end{figure}

\begin{figure}
\begin{center}
\includegraphics*[width=83mm]{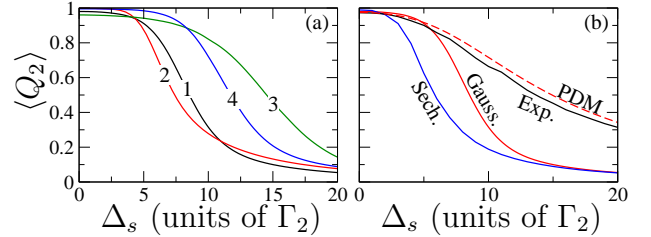}% Here is how to import EPS art
\caption{\label{linerand2} (Color online) Single resonance driven by stochastic pulses. Stochastic Auger signal averaged over 1000 stochastic pulses as a function of the detuning $\Delta_s$ for 
(a) Gaussian-correlated noise and  various combinations of pulse durations and coherence times; (b) various types of fluctuations and $\tau_s=3\Gamma_2^{-1}$, $\sigma_\omega=10\tau_s^{-1}$.  
The curves of (a) correspond to the combinations (from 1 to 4): $\tau_s=3\Gamma_2^{-1}$ and $\sigma_\omega=10\tau_s^{-1}$; $\tau_s=4.5\Gamma_2^{-1}$ and $\sigma_\omega=10\tau_s^{-1}$; 
$\tau_s=3\Gamma_2^{-1}$ and $\sigma_\omega=20\tau_s^{-1}$; $\tau_s=4.5\Gamma_2^{-1}$ and $\sigma_\omega=20\tau_s^{-1}$, respectively.
The degrees of first-order coherence for the fluctuations chosen in (b) are shown in table \ref{tab1}.  
For the sake of comparison, in Fig. \ref{linerand2} we also plot the signal corresponding 
to a phase-diffusion model (PDM) \cite{AgaPRA78,GeoPRA78} with $\gamma=2\sqrt{2\ln(2)} \sigma_\omega$. Other parameters: Gaussian profile for $\aver{I_s(t)}$, $\Omega_s^{(0)}=2\Gamma_2$,  $T_f=32\Gamma_2^{-1}$.}
\end{center}
\end{figure}

%%%%%%%
% TLA - Numerical Results
%%%%%%%

\subsection{Numerical results}
\label{sec3b}

The behaviour of $Q_2(\Delta_s)$ for a Gaussian Fourier-limited pulse of various durations and various peak Rabi frequencies is depicted in Fig. \ref{linedet}.  As depicted in Fig. \ref{linedet}(a), for fixed $\tau_s$ and $\Gamma_2$, $Q_2(\Delta_s)$ becomes wider with increasing peak Rabi frequency $\Omega_s^{(0)}$. The derivation of an analytic expression for $Q_2$ is a rather difficult task because of the time dependence of $\Omega_s(t)$. Nevertheless, our simulations show that for weak fields (small $\Omega_s^{(0)}$), the dependence $Q_2(\Delta_s)$ is well approximated by a Lorentzian. This is the case of the experimental setup in \cite{MazJPB12}. For increasing $\Omega_s^{(0)}$, however, we observe here severe deviations from the Lorentzian, especially around the peak of $Q_2(\Delta_s)$ [see Figs. \ref{linedet}(a,c)]. Furthermore, according to Figs. \ref{linedet}(b,d), $Q_s(\Delta_s)$ becomes more flat at the apex, as we increase the pulse duration $\tau_s$, for fixed $\Omega_s^{(0)}$ and $\Gamma_2$. The atoms experience large intensity for longer periods as we increase $\tau_s$, and thus the deviations from the Lorentzian extend to even larger detunings, whereas the Lorentzian seems to remain a good approximation for large $\Delta_s$ only. These observations do not depend on the actual pulse profile under consideration, and can be attributed to power broadening under the constraint of $Q_2(\Delta_s)\leq 1$ for all values of $\Delta_s$.   
Note, however, that in view of the pulsed driving, the power broadening depends on the peak intensity as well as the duration of the pulse. 

In Fig. \ref{linerand}, we show results for the average stochastic signal $\aver{Q_2}$ as a function of $\Delta_s$, in the presence of fluctuations. Recall here that the bandwidth of the stochastic pulses with Gaussian average intensity profile exceeds the bandwidth of a Fourier limited pulse of the same profile and duration by $\sqrt{1+2\chi^2}$. 
The duration of the average intensity profile has been chosen $\tau_s=3\Gamma_2^{-1}$ and the peak Rabi frequency $\Omega_s^{(0)}\leq 2\Gamma_2$. According to Fig. \ref{linedet}, the deviations from the Lorentzian shape are negligible for a Fourier-limited pulse of the same parameters. As depicted in Fig. \ref{linerand}, however, the presence of amplitude fluctuations in the pulse enhance the power broadening and thus, strong deviations from the Lorentzian appear even for the particular combinations of $\tau_s$  and $\Omega_s^{(0)}$. For  $\chi \sim 1$ the coherence time of the noise is comparable to the duration of the pulse [see Fig. \ref{linerand}(a)], and $\aver{Q_s(\Delta_s)}$ is very close to the signal for a Fourier-limited pulse with the same profile as $\aver{I(t)}$. Increasing $\chi$, we essentially decrease the coherence time relative to $\tau_s$ thus approaching the limit of stationary source ($\Delta\omega_s\approx \gamma$). Accordingly, the average stochastic signal $\aver{Q_2}$ becomes significantly broader than the signal for a Fourier-limited pulse of the same parameters, and only for large values of $\Delta_s$, the two signals approach each other [compare the dashed thin curves to the corresponding thick coloured curves in Figs. \ref{linerand}(b-d)].  Note also the prominent flattening of $\aver{Q_2}$ at the apex for $\chi\gg 1$ [see Figs. \ref{linerand}(c,d)].

Figure \ref{linerand2}(a) shows the average stochastic signal as a function of $\Delta_s$, for various combinations of $\tau_s$ and $\sigma_\omega$. Increasing the duration $\tau_s$, while keeping $\chi$ constant, results in a faster drop of $\aver{Q_2(\Delta_s)}$ as we move away from $\Delta_s=0$, and thus to a narrower $Q_2(\Delta_s)$ (compare curve 1 to curve 2, and curve 3 to curve 4). On the other hand,  if we increase $\chi$ for fixed $\tau_s$, (compare curve 1 to curve 3, and curve 2 to curve 4), the effect is the opposite. 
In Fig. \ref{linerand2}(b), we plot $\aver{Q_2(\Delta_s)}$ for fields with different types of noises (the corresponding PSDs and $g^{(1)}(t-t^\prime)$ can be found in table \ref{tab1}), but with the same $\sigma_\omega$. Clearly, in all cases $\aver{Q_2}$ flattens in the neighbourhood of $\Delta_s=0$, and the type of correlations affects mainly the wings of $\aver{Q_2(\Delta_s)}$, as we move away from $\Delta_s=0$. The faster $g^{(1)}(t-t^\prime)$ drops, the slower $\aver{Q_2(\Delta_s)}$ drops with increasing $\Delta_s$. 
Moreover, for a field with phase fluctuations only, $\aver{Q_2(\Delta_s) }$ drops much slower than for a stochastic field of the same bandwidth with both amplitude and phase fluctuations (compare PDM to Gauss. curves). 

The above discussion and findings hold for all the different profiles we used for $\langle I(t) \rangle$ throughout 
our simulations, albeit perhaps with some qualitative differences. We emphasize once more that the above discussion on the power broadening and the non-Lorentzian signals pertain to the total Auger yield as a function of the detuning of the driving FEL radiation from resonance. To the best of our knowledge, no related experimental observations have been reported in the literature so far. Perhaps the reason is that most of the related experiments have been performed in the regime of weak intensities. Our theory suggests that the required Rabi frequencies should be at least twice the natural linewidth $\Gamma_2$ of the observed Auger resonance. For the 3d$\to$5p resonance of Kr, this 
corresponds to intensities of the order of  $10^{17}$ W/cm$^2$. Such intensities are certainly within reach of various FEL facilities such  as FLASH and LCLS.

\begin{figure}
\begin{center}
\includegraphics*[width=80mm]{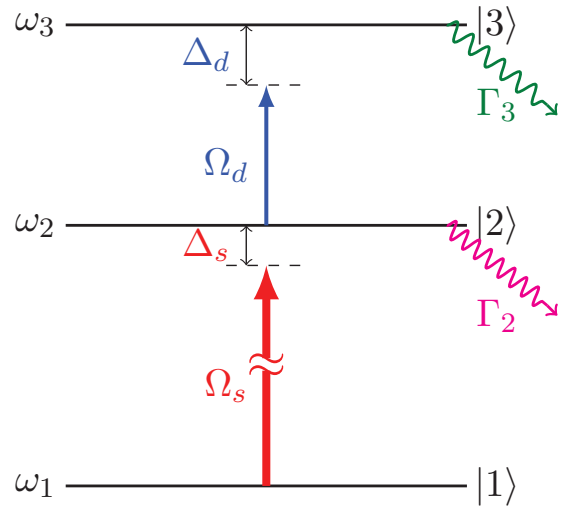}% Here is how to import EPS art
\caption{\label{Dor} (Color online) Double Auger resonance driven by a short-wavelength SASE-FEL pulse (lower transition), and a Fourier-limited pulse (upper transition). The sub-valence vacancy in the excited atom decays via two distinct Auger channels with rates $\Gamma_2$ and $\Gamma_3$.}
\end{center}
\end{figure}

%%%%%%%
% DOUBLE RESONANCE
%%%%%%%

\section{Double Auger resonance}
\label{sec4}
Consider now the case depicted in Fig. \ref{Dor}, where the core-excited state $\ket{2}$ not only decays, but it is also coupled to another state $\ket{3}$, via second field  ${\bf E}_d(t)$.

\subsection{Formalism}
In contrast to ${\bf E}_s(t)$, we assume that ${\bf E}_d(t)$ is a Fourier-limited pulse (typically of a longer wavelength) given by 
\bea
{\bf E}_d(t)=\left [{\cal E}_d(t)e^{i\omega_dt}+{\cal E}_d^\star (t)e^{-i\omega_dt}\right ]{\bf e}_d,
\eea
where $\omega_{d}$ denotes the frequency and ${\bf e}_s$ is the polarization vector. The complex amplitude ${\cal E}_d(t)$ can be written as ${\cal E}_d=\sqrt{I_d^{(0)}f_d(t)}$, where $f_d(t)$ is a Fourier-limited envelope of peak value 1, $I_d^{(0)}$ is the the peak intensity of the pulse, and let $\tau_d$ denote the corresponding duration. The transition $\ket{2}\leftrightarrow\ket{3}$ is assumed to be electric-dipole allowed, and in analogy to the definitions of the previous section, the corresponding Rabi frequency  is given by  
\bea
\Omega_d(t) = \frac{\mu_{23}\sqrt{I_d(t)}}{\hbar}= \Omega_d^{(0)}  \sqrt{f_d(t)},
\label{detOmega1}
\eea
where 
\bea
\Omega_d^{(0)} =\frac{\mu_{23}\sqrt{I_d^{(0)}}}{\hbar},
\label{detOmega2}
\eea
is the peak Rabi frequency,  and $\vec{\mu}_{23}$ the relevant transition dipole moment. The corresponding detuning from resonance is $\Delta_d=\omega_{32}-\omega_{d}$, while we must also include the Auger decay from state $\ket{3}$ with rate $\Gamma_3$. This is the so-called double-resonance (DR) setup which allows for the observation of ac Stark (Autler-Townes) splitting, and it is also employed in various quantum optics phenomena, such as the EIT \cite{Eit}.

The interaction of the atomic system with the radiation can be described by means of the reduced atomic density matrix with elements $\rho_{ij}(t)$.   
In the rotating-wave approximation, the equations of motion for $\rho_{ij}$ read  
\begin{subequations}
\label{eom2}
\bea
&&\frac{\partial\sigma_{11}}{\partial t}=2{\rm Im}[\Omega_s^\star\sigma_{12}]\\
&&\frac{\partial\sigma_{22}}{\partial t}=-\Gamma_2\sigma_{22}-2{\rm Im}\left [\Omega_s^\star\sigma_{12}-\Omega_d^\star\sigma_{23}\right ]\\
&&\frac{\partial\sigma_{33}}{\partial t}=-\Gamma_3\sigma_{33}-2{\rm Im}\left [\Omega_d^\star\sigma_{23}\right ]\\
&&\frac{\partial\sigma_{12}}{\partial t}=(i\Delta_s-\Gamma_{12})\sigma_{12}+i\left [ \Omega_s(\sigma_{22}-\sigma_{11})-\Omega_d^\star\sigma_{13}\right ]\\
&&\frac{\partial\sigma_{23}}{\partial t}=(i\Delta_d-\Gamma_{23})\sigma_{23}+i\left [ \Omega_s^\star\sigma_{13}+\Omega_d(\sigma_{33}-\sigma_{22})\right ]\\
&&\frac{\partial\sigma_{13}}{\partial t}=[i(\Delta_s+\Delta_d)-\Gamma_{13})]\sigma_{13}+i\left ( \Omega_s\sigma_{23}-\Omega_d\sigma_{12}\right ).
\eea
where $\rho_{ii}=\sigma_{ii}$, $\rho_{12}=\sigma_{12}e^{i\omega_s t}$, $\rho_{23}=\sigma_{23}e^{i\omega_d t}$, $\rho_{13}=\sigma_{13}e^{i(\omega_s+\omega_d) t}$.  
In  the absence of other types of (in)homogeneous broadening mechanisms the off-diagonal relaxation rates are $\Gamma_{ij}=(\Gamma_i+\Gamma_j)/2$.
 
The frequency response of the atoms is obtained by monitoring the Auger electrons. The total probability for Auger decay from state $\ket{j}$ is given by 
\bea
Q_j = \Gamma_j\int_0^{\infty} \sigma_{jj}(t)\,{\rm d}t.
\eea 
\end{subequations}
In each realization of our simulations a SASE-FEL pulse ${\cal E}_s(t)$ is generated randomly, according to the algorithm described in Sec. \ref{sec2}, and the above set of stochastic differential equations is propagated from $t=0$ to $t=T_f$ with $T_f\gg\max\{\tau_s,\tau_d\}$, keeping track of $Q_j$ at $t=T_f$. The average stochastic signal $\aver{Q_j}$ is obtained by averaging over a large number of  pulses.  

One can distinguish between two different arrangements of weak  and strong  pulses in the context of DR. In the first arrangement, the Fourier-limited pulse is strong (pump) and the stochastic SASE-FEL pulse is weak, and is used as a probe for monitoring the upper transition. This arrangement is pertinent to EIT and has been discussed recently for optical control of X-ray absorption \cite{Eit}. Alternatively, one may also consider a strong stochastic pump in the lower transition, and a weak Fourier-limited probe in the upper one. We have explored both arrangements and in the following we present our main findings on the effect of fluctuations in the DR setup. To facilitate the characterization of such effects, in the appendix we summarize briefly some of the fundamental aspects of the problem in the absence of fluctuations (Fourier-limited pulses).

\subsection{Numerical results}
In this section we discuss the effects of fluctuations in the DR signal. The  combinations chosen for peak Rabi frequencies and durations for 
the pump/probe pulse profiles are such that, in the case of Fourier-limited pulses, $Q_j$ exhibits a clear doublet when plotted as a function of the detuning of the probe from resonance.  The deterministic signal $Q_j$ for Fourier-limited pulses serves as a reference for the investigation of effects induced by the fluctuations in the probe and the pump. In other words, the average stochastic signal $\aver{Q_j}$ in the presence of fluctuations, is compared to the deterministic signal $Q_j$ pertaining to Fourier-limited pump/probe 
pulses. 

\begin{figure}
\begin{center}
\includegraphics*[width=85mm]{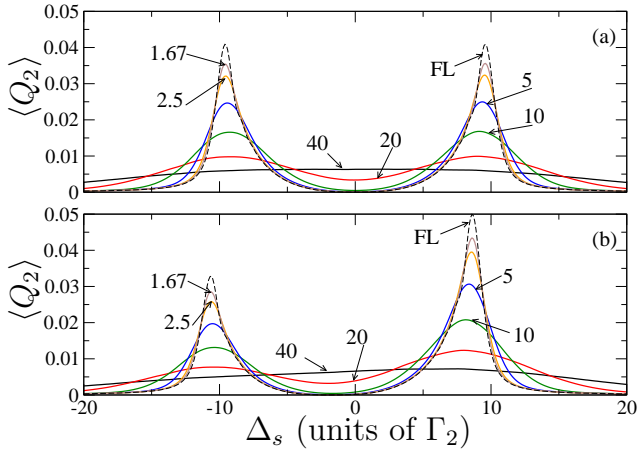}% Here is how to import EPS art
\caption{\label{dor-st1} (Color online) DR arrangement I with stochastic probe  and Fourier-limited pump.  The average stochastic signal is plotted as a function of the probe detuning  $\Delta_s$  for increasing values of  $\chi$, and for (a) resonant ($\Delta_d=0$) and (b) off-resonant ($\Delta_d=2\Gamma_2$)  pump.  The thin dashed curves correspond to a Fourier limited probe (i.e., $\chi\ll 1$). Synchronized  Gaussian profiles have been chosen for the average probe intensity $\aver{I_s(t)}$ as well as  for the pump. Other parameters: $\tau_s=4.5\Gamma_2^{-1}$,  $\tau_d=6\Gamma_2^{-1}$, $\Omega_s^{(0)}=0.1\Gamma_2$, $\Omega_d^{(0)}=10\Gamma_2$, $\Gamma_3=\Gamma_2$, $T_f=32\Gamma_2^{-1}$, 5000 realizations, Gaussian-correlated noise.}
\end{center}
\end{figure}

\begin{figure}
\begin{center}
\includegraphics*[width=85mm]{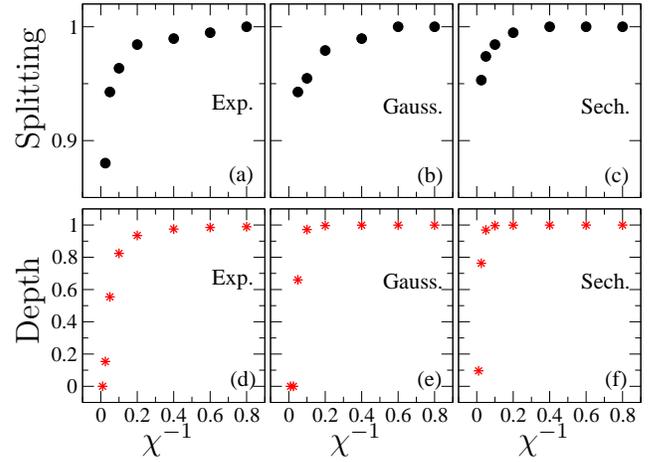}% Here is how to import EPS art
\caption{\label{sd-fig} (Color online) DR arrangement I with stochastic probe  and Fourier-limited pump. (a-c) The separation of the peaks in the average stochastic signal $\aver{Q_2}$, normalized to the separation in the signal for Fourier-limited probe of the same shape (approximately equal to $19.2\Gamma_2$), as a function of $\chi^{-1}$ for various noises. (d-f) The depth of the doublet in the average stochastic signal 
(given by Eq. \ref{depth}),  as a function of $\chi^{-1}$ for various noises.  Other parameters as in Fig. \ref{dor-st1}.} 
\end{center}
\end{figure}

\subsubsection{Arrangement I: stochastic probe} 
Although usually one would want the probe not to have fluctuations, in considering FEL as a probe we are compelled to  consider the effect of fluctuations (e.g., this is the case of the experiment \cite{Eit}). 
The stochastic probe is applied on $\ket{1}\leftrightarrow\ket{2}$, and a Fourier-limited pump on $\ket{2}\leftrightarrow\ket{3}$. The average stochastic signal $\aver{Q_2}$, is plotted as a function of $\Delta_s$ in Fig. \ref{dor-st1}, for various values of $\chi$ and two different detunings of the pump $\Delta_d$. For the sake of comparison, the deterministic signal for Fourier-limited probe pulse with profile $\aver{I_s(t)}$ is also shown (dashed curve). For $\chi\sim 1$, the signal is very close to the one for Fourier-limited probe. 
As we increase $\chi$ (i.e., as we decrease $T_c\sim\sigma_\omega^{-1}$ relative to $\tau_s$), the two peaks of the doublet become broader, whereas their position is also affected slightly. Recall here that according to Eq. (\ref{bw1c}), $\Delta\omega_s\approx \gamma$ for $\chi>2$, with the bandwidth $\gamma$ given in the second row of table \ref{tab1}.
Thus, the visibility of the doublet decreases with increasing bandwidth (or decreasing $T_c$), and for $\chi>20$ it has  essentially disappeared. It is worth noting here that this critical 
value of $\chi$ above which no doublet is observed corresponds to $\gamma\approx \Omega_d$. In the picture of eigenstates $\ket{\pm}$ (see appendix), the fluctuating probe leads to fluctuating  couplings $|\langle 1|\pm\rangle|$, which in turn give rise to a noisy signal, i.e., to broader peaks. When this broadening exceeds the peak Rabi  frequency of the pump, the doublet disappears.

In Figs. \ref{sd-fig}(a-c), we plot the separation of the peaks in the doublet as a function of $\chi^{-1}$, for various types of noises (see table \ref{tab1}). For the sake of comparison, the separation is normalized to its value for Fourier-limited probe with the same profile as $\aver{I(t)}$ (see dashed curve in Fig. \ref{dor-st1}).  As was expected, in all cases the normalized splitting approaches 1 as $\chi\to 1$, and decreases as we decrease $\chi^{-1}$ (i.e., as we decrease $T_c\sim\sigma_\omega^{-1}$ for fixed $\tau_s$). The decrease depends weakly on the type of noise in the probe, and for $\chi^{-1}\geq 0.05$ it does not exceed 6\%.  In this sense, the position of the peaks is not affected significantly by the presence of  fluctuations in the probe, and this enables us to deduce the dipole moment for the transition between the Auger states $\ket{2}$ and $\ket{3}$. Indeed, the pump is Fourier limited and in principle one can have control on the peak intensity $I_d^{(0)}$. Hence, in view of Eq. (\ref{detOmega2}), one can deduce $\mu_{23}$ by estimating $\Omega_d^{(0)}$ through the separation of the peaks in the average stochastic signal $\aver{Q_2(\Delta_s)}$,  for $\Delta_d=0$. As discussed in the appendix, for Fourier-limited probe and pump pulses, the separation of the peaks is  not precisely equal to $2\Omega_d^{(0)}$, but it is very close [$\simeq 8\%$ smaller for the parameters of Fig. \ref{dor-st1}(a)]. Thus, obtaining  $\mu_{23}$ from the splitting in $\aver{Q_2(\Delta_s)}$ will unavoidably lead to an error due to the time-dependence of the pump and the probe, as well as an error due to the fluctuations in the probe. Assuming that the Fourier-limited profile of the pump is known, the former error could be eliminated only if one can measure the average intensity profile for the probe $\aver{I_s(t)}$ together with $\aver{Q_2(\Delta_s)}$. In any case, we believe that for judiciously chosen probe/pump pulse durations, and coherence time for the probe, this approach allows one to measure the  dipole moment between Auger states with at most $20\%$ of error. To the best of our knowledge, this is the only way to the measurement of the dipole moment between two Auger or autoionizing states, which provides a very sensitive test of theoretical models for states embedded in continua.

Besides the splitting, the visibility of the doublet also depends on its depth which can be quantified by
\bea \label{depth} V = \frac{\aver{Q_j(\Delta_s)}_{\max}-\aver{Q_j(\Delta_s)}_{\min}}{\aver{Q_j(\Delta_s)}_{\max}},\eea
and is also plotted in Figs. \ref{sd-fig}(d-f) as a function of $\chi^{-1}$ for various noises. For $\chi\approx 1$, $V$ is close to $1$ and decreases with decreasing $\chi^{-1}$. The critical value of $\chi^{-1}$ below which we observe a rapid decrease of the depth is approximately $0.1$, and depends weakly on the type of the noise in the probe. It basically corresponds to FEL bandwidth $\gamma\approx \Omega_d/2$. As a general remark, note that there is a rather large regime of values for $\chi$, where the depth and the splitting do not vary appreciably. Hence, none of them seems capable of resolving reliably different values of $\chi$, providing thus information about the actual bandwidth of the probe. The question arising therefore is whether the width of the peaks are more sensitive to variations of $\chi$, and may thus provide more information in this respect. 

\begin{figure}
\begin{center}
\includegraphics*[width=85mm]{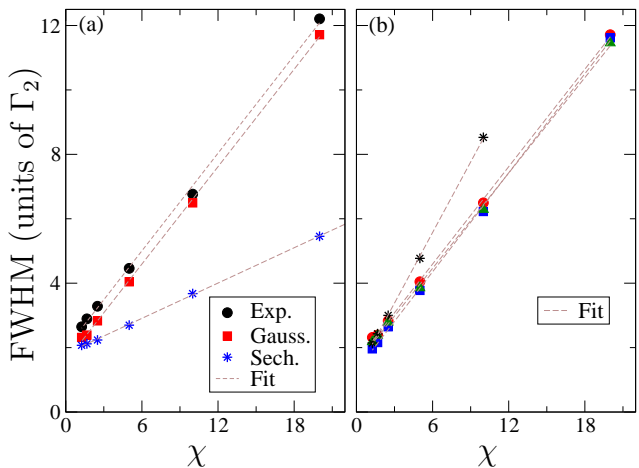}% Here is how to import EPS art
\caption{\label{fwhm.fig} (Color online) DR arrangement I with stochastic probe and Fourier-limited pump. (a) The FWHM of the two peaks in the average stochastic signal  $\aver{Q_2(\Delta_s)}$, is plotted as a function of $\chi$, for various types of noises. Other parameters as in Fig. \ref{dor-st1}. (b) As in Fig. \ref{fwhm.fig}(a), but for Gaussian correlated noise only, and various combinations of atomic and field parameters. (Circle) Gaussian profile for $\aver{I_s(t)}$, $\tau_s=4.5\Gamma_2^{-1}$, $\Gamma_3=\Gamma_2$; (Square) Gaussian profile for $\aver{I_s(t)}$, $\tau_s=4.5\Gamma_2^{-1}$,  $\Gamma_3=0.5\Gamma_2$; (Star) Gaussian profile for $\aver{I_s(t)}$, $\tau_s = 3\Gamma_2^{-1}$, $\Gamma_3=\Gamma_2$; 
(Triangle) profile 2 for $\aver{I_s(t)}$, $\tau_s=4.5\Gamma_2^{-1}$, $\Gamma_3=\Gamma_2$.  Other parameters: Gaussian pump pulse, $\tau_d=6\Gamma_2^{-1}$, $\Omega_s^{(0)}=0.1\Gamma_2$, $\Omega_d^{(0)}=10\Gamma_2$, $T_f=32\Gamma_2^{-1}$, 5000 realizations.}
\end{center}
\end{figure}

In Fig.  \ref{fwhm.fig}(a) we plot the mean value of the FWHM for the two peaks in the average stochastic signal, as a function of $\chi$, for various types of noises in the probe pulse. Clearly in all of the cases, we have a linear dependence on $\chi$. The corresponding slope is the same for Gaussian and exponentially correlated noise, whereas it is smaller for noise with correlations that drop as a hyperbolic secant (see table \ref{tab1}). According to Fig. \ref{fwhm.fig}(b),  the linear dependence and the slope depend only weakly on the 
 profile of the average probe intensity $\aver{I_s(t)}$ and the decay rate of state $\ket{3}$. However, the slope does seem to have an inversely proportional dependence on the duration of  $\aver{I_s(t)}$ (this was to be expected of course, in view of the discussion in the appendix). These observations suggest a strong correlation between the FWHM  of the peaks 
in the average stochastic signal and the bandwidth of the probe $\Delta\omega_s$, from the regime of Fourier-limited probes ($\chi\ll 1$) all the way up to stationary probes ($\chi\gg 1$). 

\subsubsection{Arrangement II: stochastic pump}
\begin{figure}
\includegraphics*[width=85mm]{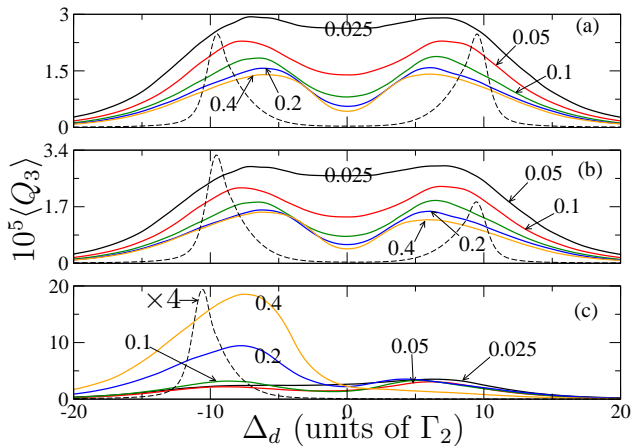}% Here is how to import EPS art
\caption{\label{dor-st3} (Color online) DR arrangement II with stochastic pump  and Fourier-limited probe. The average stochastic signal is plotted as a function of the probe 
detuning $\Delta_d$, for increasing values of the ratio $\chi$, and for pump detunings (a)  $\Delta_s=0$; (b) $\Delta_s=0.1\Gamma_2$; (c) $\Delta_s=2\Gamma_2$. The thin dashed curves correspond to a Fourier-limited pump (i.e., $\chi\ll 1$). Synchronized  Gaussian profiles have been chosen for the average pump intensity $\aver{I_s(t)}$ as well as  for the probe. Other parameters: $\tau_s=6\Gamma_2^{-1}$,  $\tau_d=3\Gamma_2^{-1}$, $\Omega_s^{(0)}=10\Gamma_2$, $\Omega_d^{(0)}=0.1\Gamma_2$, $\Gamma_3=\Gamma_2$, $T_f=32\Gamma_2^{-1}$, Gaussian-correlated noise.}
\end{figure}

For this arrangement, the transition $\ket{1}\leftrightarrow\ket{2}$ is strongly driven by stochastic pulses, whereas the Fourier-limited probe is applied on $\ket{2}\leftrightarrow\ket{3}$. The splitting induced in the lower transition is monitored by keeping track of the Auger electrons from the decay of state $\ket{3}$. 
The average stochastic signal is depicted in Fig. \ref{dor-st3}, as a function of the probe detuning $\Delta_d$, and for three different detunings of the pump. Clearly, the stochastic pump has a dramatic effect on
the doublet, as compared to the case of Fourier-limited pump (dashed curves). The peaks of the doublet broaden significantly even for values of $T_c$ comparable to $\tau_s$, whereas at the same time their separation is reduced.  

The effect of fluctuations in this arrangement are far more severe than in arrangement I, because they pertain to the pump pulse, and not to the probe. The atoms experience different intensities at different times within a single pump pulse, and the induced splitting also changes with time. In the picture of eigenstates (see appendix), instead of two states $\ket{\pm}$ with well-defined energies, one has essentially two manifolds of $\ket{+}$ and $\ket{-}$ states, and there are no clear resonances for the probe to resolve. Hence, the distribution of $Q_3(\Delta_d)$ 
exhibits a  rather strong background and the visibility of the doublet is significantly reduced.  The background becomes stronger as we increase $\chi$ for fixed $\tau_s$ (i..e, with increasing bandwidth or decreasing $T_c$), leading to severe qualitative and not only quantitative deviations from the corresponding signal for Fourier-limited pump.  By contrast to arrangement I, 
in the present arrangement the width of the peaks in the doublet are comparable to  their separation, which reduces even further the visibility of the doublet. Hence, it is rather difficult to 
extract from the doublet, accurate information e.g., about the bandwidth of the FEL pulses. In this respect, therefore the previous arrangement is far more useful for spectroscopic applications.

\section{Summary}
\label{sec5}
We have discussed the effects of field fluctuations on the frequency response of single and double Auger resonances driven by FEL radiation. We have found that, in the case of single resonance the fluctuations tend to enhance power broadening phenomena, leading thus to unconventional distributions for the total yield of Auger electrons with respect to the detuning of the driving field from resonance.  In the case of double resonance, when 
the FEL radiation plays the role of the probe, the total yield of the Auger electrons exhibits the characteristic Autler-Townes doublet when the bandwidth of the source does not exceed the 
Rabi frequency of the pump. The widths of the peaks have been shown to be strongly correlated  with the bandwidth of the FEL pulses, whereas their separation allows for the estimation 
of the dipole moment between the Auger resonances involved in the double-resonance setup. Finally, when FEL radiation is used to pump the double-resonance, field fluctuations have been shown to have a dramatic effect on the ac Stark splitting.  The modelling of fluctuations and the parameters used throughout this work follow closely real experimental situations, and thus our findings can be tested in practise.

Although our discussion and quantitative illustration of the physics has been in the context of Auger resonances, the formalism and treatment are equally valid for any type of decaying resonances, in any system. In closing, we should mention that we are well aware of the effects of interaction-volume expansion in any situation involving strong radiation, which by necessity is focused. This is an instrumental effect which is apt to affect the observed line shapes, and as such needs to be taken into consideration in the interpretation of experimental data. It does, however, depend on the particular focusing geometry pertaining to a given experiment, but the relevant theoretical tools are known \cite{LamPRA11}. In any case, having results without the volume effect provides a point of calibration to be taken into consideration in the estimation of its expected influence.

\section*{Acknowledgements}
The authors acknowledge with pleasure discussions at various times with R. Santra and N. Rohringer, on issues pertaining to field fluctuations.

\appendix*
\section{DR with Fourier-limited pulses}
\label{sec4a}
The problem of DR with Fourier-limited pulses has been studied extensively in the past, and the reader may refer to related literature (e.g., see \cite{NikPRA11,RodJMO87,LauOA86,WhiPRA76,GreJPB85,FerPRA05,Shore} and references therein). To facilitate the discussion of Sec. \ref{sec4}, we summarize briefly here fundamental aspects of this problem.

\begin{figure}
\begin{center}
\includegraphics*[width=80mm]{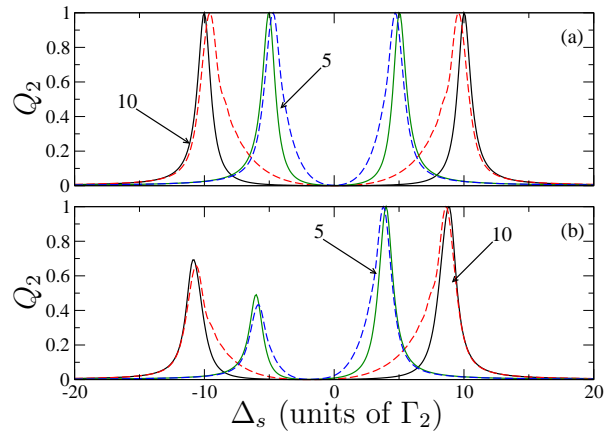}% Here is how to import EPS art
\caption{\label{det0} (Color online) DR arrangement I  with Fourier-limited pump and probe.  The Auger signal $Q_2$ is plotted as a function of the detuning $\Delta_s$, for two different ratios of $\Omega_d^{(0)}/\Gamma_2$ and pump detunings (a) $\Delta_d=0$; (b) $\Delta_d=2\Gamma_2$. 
The dashed curves correspond to synchronized Gaussian probe and pump pulses with durations $\tau_s=4.5\Gamma_2^{-1}$ and  $\tau_d=6\Gamma_2^{-1}$, respectively. The solid curves correspond to constant pump and probe fields lasting for time equal to  $T_f$. For the sake of comparison, the signal has been normalized to its maximum value. Other parameters: $\Omega_s^{(0)}=0.1\Gamma_2$, $\Gamma_3=\Gamma_2$, $T_f=32\Gamma_2^{-1}$.  }
\end{center}
\end{figure}

\begin{figure}
\begin{center}
\includegraphics*[width=80mm]{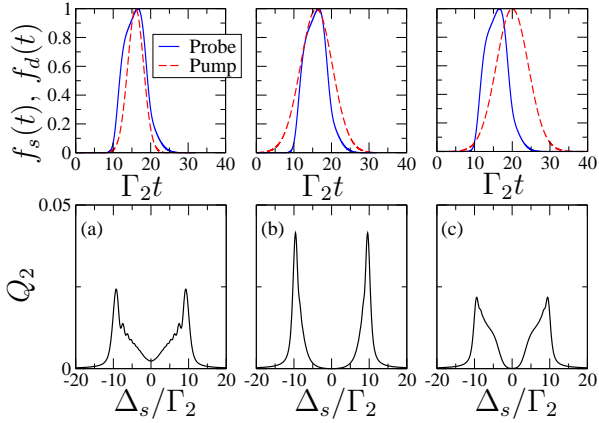}% Here is how to import EPS art
\caption{\label{det2} (Color online) DR arrangement I  with Fourier-limited pump and probe. (a-c) Signal $Q_2$ at the end of the sequence of probe/pump pulses shown in the upper panel, as a function of the detuning $\Delta_s$. 
The probe follows the profile 2, discussed in Sec. \ref{sec2}, and its FWHM is approximately $7.5\Gamma_2^{-1}$. The pump pulse is taken Gaussian with duration: (a) $\tau_d=3\Gamma_2^{-1}$;  (b,c) $\tau_d=6\Gamma_2^{-1}$. 
For (a) and (b) the probe is delayed relative to the pump by approximately $0.6\Gamma_2^{-1}$, whereas for (c) the probe precedes by approximately $4\Gamma_2^{-1}$.  
Other parameters: $\Omega_s^{(0)}=0.1\Gamma_2$, $\Omega_d^{(0)}=10\Gamma_2$, $\Gamma_3=\Gamma_2$, $\Delta_d=0$, $T_f=32\Gamma_2^{-1}$. }
\end{center}
\end{figure}

\begin{figure}
\begin{center}
\includegraphics*[width=80mm]{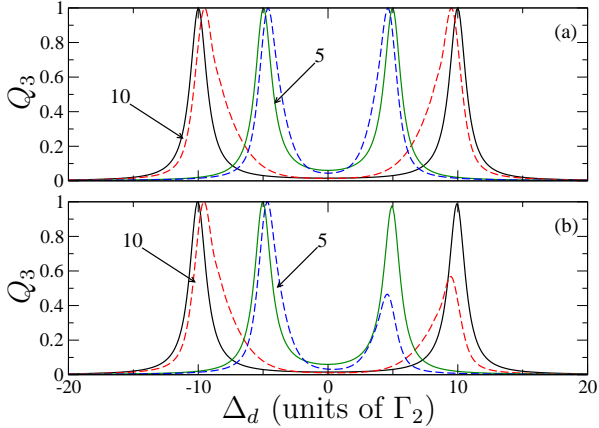}% Here is how to import EPS art
\caption{\label{det3} (Color online)  DR arrangement II with Fourier-limited pump and probe. The Auger Signal $Q_3$ is plotted as a function of the detuning $\Delta_d$, for two different ratios of $\Omega_s^{(0)}/\Gamma_2$, and for pump detunings (a) $\Delta_s=0$; (b) $\Delta_s=0.1\Gamma_2$. 
The dashed curves correspond to Gaussian probe and pump pulses with durations $\tau_d=3\Gamma_2^{-1}$ and  $\tau_s=6\Gamma_2^{-1}$, respectively. The solid curves correspond to constant pump and probe fields lasting for time equal to $T_f$. For the sake of comparison, the signal has been normalized to its maximum value. Other parameters: $\Omega_d^{(0)}=0.1\Gamma_2$, $\Gamma_3=\Gamma_2$, $T_f=32\Gamma_2^{-1}$.}
\end{center}
\end{figure}

\subsection{Arrangement I: probe in the lower transition}
Consider the case where the upper transition is strongly driven, and the lower one is weakly driven by the probe. The strong driving of the upper transition results in a 
splitting referred to as  ac Stark splitting. In the case of stationary fields (i.e., for time independent ${\cal E}_{s(d)}(t)$), one can readily show this splitting by diagonalizing the Hamiltonian corresponding to the strongly-driven transition $\ket{2}\leftrightarrow\ket{3}$. There are two eigenenergies  given by 
\bea
\omega_\pm = -\frac{\Delta_d}2\pm\frac{1}2\sqrt{\Delta_d^2+4\Omega_d^2},
\eea
with the corresponding eigenstates $\ket{\pm}=\mp\sin(\theta_{\pm})\ket{2}+\cos(\theta_{\pm})\ket{3}$,
where $\tan(\theta_\pm)=\mp \Omega_d/\omega_{\pm}$. The induced ac Stark splitting is simply $\omega_+-\omega_-=\sqrt{\Delta_d^2+4\Omega_d^2}$, and can be monitored by following the absorption of a weak probe,  as we vary its frequency around the lower resonance $\ket{1}\leftrightarrow\ket{2}$. The DR setup in this picture involves  two distinct absorption paths $\ket{1}\to\ket{\pm}$ for the probe, with the ratio of the corresponding couplings determined by the overlaps i.e., $\mu\equiv|\langle 1\ket{-}|/|\langle 1\ket{+}|$. For $\Delta_d=0$, $\mu=1$ and thus the two absorption channels are equivalent, whereas for 
$\Delta_d>0$ the channel $\ket{1}\to\ket{+}$ dominates over the channel $\ket{1}\to\ket{-}$, and the opposite is true for $\Delta_d<0$. The absorption of the probe with varying $\Delta_s$ can be monitored directly (e.g., via a detector), or indirectly e.g., by examining the Auger electrons from level $\ket{2}$; larger absorption yields stronger signal. 
Such a signal in presented in Fig. \ref{det0}, for two different Rabi frequencies and pump detunings (see solid lines). Clearly, $Q_2$ exhibits a doublet with the two peaks located at $\omega_\pm$, and the highest peak being the one at $\omega_+$ for $\Delta_d> 0$. Moreover, in the regime of $\Delta_s=0$ the signal is negligible. In the spirit of EIT, this corresponds to a transparency window, where the absorption of the probe is reduced significantly to frequencies around $\omega_{21}$. 

The previous qualitative description also holds for pulsed pump and probe. In this case, however, the atoms experience different intensities at different times, and this results in a  considerable broadening of the peaks, as well as a small decrease of the separation of the peaks in the doublet (see dashed curves in Fig. \ref{det0}). The FWHM of the peaks in the doublet acquires a dependence on the the pulse duration, in addition to the dependence on the linewidths of the states involved.  In contrast to stationary fields, there are many additional parameters in the problem pertaining to the  details of the pulses (such as shapes, durations, peak values, etc) as  well as their relative position. As a rule of thumb, in order for a clear doublet to be observable, the atoms have to be ``dressed" by the pump field before the probe rises, so that the splitting has been induced. The conditions under which such a requirement is fulfilled depends on the particular pulse shapes under consideration. Roughly speaking, the duration of the pump must be sufficiently longer than the duration of the probe pulse (i.e., $\tau_d>\tau_s$), and any relative delay between the pulses should be sufficiently small so that the overlap between the pulses is not  reduced significantly. As depicted in Fig. \ref{det2}, if these conditions are fulfilled, a clear doublet can be observed irrespective of the actual profile of the probe, whereas in the opposite case the doublet may be distorted by additional artifacts (essentially the probe resolves various irrelevant details during the rise and the fall of the pump pulse). The fact 
that the observability of the doublet does not seem to depend on the  form of the pulse shapes is of particular interest for the discussion in Sec. \ref{sec4},  because when SASE-FEL pulses are involved in the DR, one cannot predict in advance the precise profile for the average intensity $\aver{I_s(t)}$.

\subsection{Arrangement II: probe in the upper transition}
In this case, the diagonalization has to be performed on the part of the  Hamiltonian that pertains to the strongly driven transition $\ket{1}\to\ket{2}$. The above expressions for the eigenergies and eigenstates hold, with the replacements $\Omega_d\to\Omega_s$, $\Delta_d\to \Delta_s$, $\ket{2}\to \ket{1}$ and $\ket{3}\to \ket{2}$.  By contrast to the previous arrangement, the absorption of the probe now involves the two distinct paths $\ket{\pm}\to\ket{3}$, and can be monitored e.g., by following  the Auger electrons  from level $\ket{3}$. The couplings for the two absorption channels  $\ket{\pm}\to\ket{3}$ depend on the overlaps $|\langle 3\ket{\pm}|$, with the corresponding ratio being equal to  $1/\mu$. As a result, in the case of off-resonant drive (i.e., for $\Delta_d\neq0$) the position of the highest peaks in the asymmetric doublet is opposite to the one in arrangement I. However, the signal $Q_3$ is substantially weaker than in the previous arrangement, because of the admixture of state $\ket{1}$ with the decaying state $\ket{2}$ in the eigenstates. The strong driving of $\ket{1}\leftrightarrow\ket{2}$ results in a fast depletion of the population by optical pumping, and thus to significant losses that do not contribute to $Q_3$. 
For the same reason, in the case of off-resonant drive, the shorter peak in the doublet seems to be suppressed severely, even for a small increase of $|\Delta_s|$. 
Finally, the effects of time-dependent pump and probe on the ac Stark splitting and its observability are analogous to the ones in arrangement I.


\begin{thebibliography}{99}
\bibitem{AgaPRA70} G. S. Agarwal, Phys. Rev. A {\bf 1}, 1445 (1970).

\bibitem{LamAAM76} P. Lambropoulos, Advances in Atomic and Molecular Physics {\bf 12}, 87 (1976).

\bibitem{ZolPRA79} P. Zoller, Phys. Rev. A {\bf 20}, 1019 (1979).

\bibitem{AgaPRA78} G. S. Agarwal, Phys. Rev. A {\bf 18}, 1490 (1978).

\bibitem{GeoPRA78} A. T. Georges and P. Lambropoulos, Phys. Rev. A {\bf 18}, 587 (1978).

\bibitem{GeoPRA79} A. T.  Georges and P. Lambropoulos, Phys. Rev. A {\bf 20}, 991 (1979). 

\bibitem{AgoJPB78} P. Agostini {\em et al.},  J. Phys. B: At. Mol. Opt. Phys. {\bf 11} 1733 (1978). 

\bibitem{CamPRA93} J. C. Camparo and P. Lambropoulos, Phys. Rev. A {\bf 47}, 480 (1993).

\bibitem{GlaPR63} R. J. Glauber, Phys. Rev. {\bf 131}, 2766 (1963).

\bibitem{ScrJPB06} A. Scrinzi {\em et al.}.  J. Phys. B: At. Mol. Opt. Phys. {\bf 39}, R1 (2006).

\bibitem{LamPRA11} P. Lambropoulos, G. M. Nikolopoulos and K. G. Papamihail, Phys. Rev. A {\bf 83}, 021407(R) (2011), and references therein.


\bibitem{LamJPB11} P. Lambropoulos, K. G. Papamihail, and P. Decleva, J. Phys. B: At. Mol. Opt. Phys. {\bf 44}, 175402 (2011).

\bibitem{MazJPB12} T. Mazza {\em et al.}, J. Phys. B: At. Mol. Opt. Phys. {\bf 45} 141001 (2012). 

\bibitem{SalSchYur} E. L. Saldin, E. A. Schneidmiller and M. V. Yurkov, Opt. Commun. {\bf 148}, 383 (1998); Nucl. Instrum. Methods A {\bf 507}, 106 (2003);   New J. Phys. {\bf 12}, 035010 (2010).

\bibitem{Kri06} S. Krinsky and Y. Li, Phys. Rev. E {\bf 73}, 066501 (2006); S. Krinsky and R. L. Gluckstern, Phys. Rev. ST Accel. Beams {\bf 6}, 050701 (2003);

\bibitem{Ack07} W. Ackermann {\em et al.}, Nature Photonics {\bf 1}, 336 (2007); P. Emma {\em et al.},  Nature Photonics {\bf 4}, 641 (2010) and references therein.

\bibitem{Eit} T.E. Glover {\em et al.}, Nature Physics {\bf  6}, 69 (2010), and references therein; C. Buth, R. Santra and L. Young, Phys. Rev. Lett. {\bf 98}, 253001 (2007), and references therein.

\bibitem{LamPRA81} P. Lambropoulos and P. Zoller, Phys. Rev. A {\bf 24}, 379 (1981).

\bibitem{KarPRL95} N. E. Karapanagioti {\em et al.}, Phys. Rev. Lett. {\bf 74}, 2431 (1995). 

\bibitem{TheJPB04} S. Themelis, P. Lambropoulos, and M. Mayer,  J. Phys. B: At. Mol. Opt. Phys. {\bf 37} 4281 (2004). 

\bibitem{LohCP08} Z.-H. Loh, C. H. Greene, S. R. Leone, Chemical Ohys. {\bf 350}, 7 (2008).

\bibitem{NikPRA11} L. A. A. Nikolopoulos, T. J. Kelly, and J. T. Costelo, Phys. Rev. A {\bf 84}, 063419 (2011).

\bibitem{RohPRA08} N. Rohringer and R. Santra, Phys. Rev. A {\bf 77}, 053404 (2008).

\bibitem{KanPRL11}  E. P. Kanter {\em et al.}, Phys. Rev. Lett. {\bf 107}, 233001 (2011). 

\bibitem{Pie96Rei99} P. Pierini, and W. Fawley, Nucl. Instrum. Methods A {\bf 375}, 332 (1996);  S. Reiche, Nucl. Instrum. Methods A {\bf 429}, 243 (1999).

\bibitem{Loud} R.~Loudon, {\em The Quantum Theory of Light}, 3rd ed. (Oxford University Press, Oxford, 2000). 

\bibitem{Good} J. W. Goodman, {\em Statistical Optics}, (Wiley, New York, 1985).


\bibitem{FoxPRA88} R. F. Fox, I. R. Gatland, R. Roy, and G. Vemuri, Phys. Rev. A {\bf 38}, 5938 (1988); R. Mannella and V. Palleschi, Phys. Rev. A {\bf 40}, 3381 (1989).

\bibitem{VanTeiAO80} G. Vannucci and M. C. Teich, Appl. Opt. {\bf 19}, 548 (1980).

\bibitem{BilShiPRA90} K. Y. R. Billah and M. Shinozuka, Phys. Rev. A {\bf 42}, 7492(R) (1990); Phys. Rev. A {\bf 46}, 8031 (1992); R. Mannella and V. Palleschi, Phys. Rev. A {\bf 46}, 8028 (1992).

\bibitem{VartPRL11} I. A. Vartanyants {\em et al.}, Phys. Rev. Lett. {\bf 107}, 144801 (2011).

\bibitem{MitzOE08} R. Mitzner {\em et al.}, Optics Express {\bf 16}, 19909 (2008).


\bibitem{Yariv} A. Yariv and P. Yeh, {\em Photonics: Optical electronics in modern communications}, (Oxford University Press, New York 2007).


\bibitem{remark1} Although we have adopted the term ``realization" throughout this work, other authors may use the equivalent term ``trajectory".

\bibitem{RodJMO87} P. A. Rodgers and S. Swain, J. Mod. Opt. {\bf 34}, 643 (1987); P. A. Rodgers and S. Swain, J. Mod. Opt. {\bf 36}, 941 (1989).

\bibitem{LauOA86} M. A. Lauder, P. L. Knight, and P. T. Greenland, Opt. Acta {\bf 33}, 1231 (1986).

\bibitem{WhiPRA76} R. M. Whitley, and C. R. Stroud, Phys. Rev. A {\bf 14}, 1498 (1976).

\bibitem{GreJPB85}  P. T. Greenland, J. Phys. B {\bf 18}, 401 (1985).

\bibitem{FerPRA05} R. Garcia-Fernadez {\em et al.}, Phys. Rev. A {\bf 71}, 023401 (2005).

\bibitem{Shore} B. W. Shore, {\em The theory of Coherent Atomic Excitation}, (Wiley, New York, 1990).



\end{thebibliography}
\end{document}